\documentclass[11pt]{article}
\usepackage{amsmath,amssymb}
\usepackage{hyperref}
\usepackage{color}
\usepackage{relsize}
\usepackage{yfonts}
\usepackage{amsbsy}
\usepackage{graphicx}  

\usepackage{mathtools}
\usepackage{makecell}

\usepackage[T1]{fontenc}

\newsavebox{\uuunit}
\sbox{\uuunit}
    {\setlength{\unitlength}{0.825em}
     \begin{picture}(0.6,0.7)
        \thinlines
        \put(0,0){\line(1,0){0.5}}
        \put(0.15,0){\line(0,1){0.7}}
        \put(0.35,0){\line(0,1){0.8}}
       \multiput(0.3,0.8)(-0.04,-0.02){10}{\rule{0.5pt}{0.5pt}}
     \end {picture}}

\newcommand{\ba}{\begin{eqnarray*}}
\newcommand{\ea}{\end{eqnarray*}}
\newcommand{\ban}{\begin{eqnarray}}
\newcommand{\ean}{\end{eqnarray}}

\def\beq{\begin{equation}}
\def\bee{\begin{equation}}
\def\eeq{\end{equation}}
\def\bea{\begin{eqnarray}}
\def\eea{\end{eqnarray}}
\def\bd{\begin{displaymath}}
\def\ed{\end{displaymath}}

\usepackage{tikz}

\numberwithin{equation}{section}
\begin{document}

%nice sentence in Louis et al
%The existence of moduli spaces in supersymmetric compactifications of string theory to four dimensions 
%impedes the construction of realistic models of particle physics and cosmol- ogy, and a primary endeavor in 
%string phenomenology is the study of mechanisms that lift the vacuum degeneracy and stabilize the moduli.

\thispagestyle{empty}
{}

%\hfill

\vskip -3mm
\begin{center}
{\bf\LARGE
\vskip - 1cm
  Quantum Gates  to other   Universes \\ [2mm]}

\vspace{15mm}

{\large
{\bf Constantin Bachas \  and\    Ioannis Lavdas }
 
\vspace{1cm}

{  Laboratoire de Physique Th\'eorique de l' \'Ecole Normale
Sup\'erieure,\\
PSL Research University, CNRS, Sorbonne Universit\'es, UPMC
Univ.\,Paris 06,\\
24 rue Lhomond, 75231 Paris Cedex 05, France  \\ 
 
}}

\vspace{5mm}

\end{center}
\vspace{5mm}

\begin{center}
{\bf ABSTRACT}\\
\end{center}
\smallskip
{We present   a  microscopic model of  a  bridge connecting   two large  Anti-de-Sitter  Universes. 
The  Universes admit  a holographic description as three-dimensional ${\cal N}=4$ supersymmetric
gauge theories
based on   large linear quivers, and   the bridge is a small rank-$n$    gauge group that
acts as a messenger.
On  the gravity side,    the bridge is a  piece of a highly-curved   AdS$_5\times$S$^5$  throat
carrying   $n$ units of five-form flux. 
We derive a
 universal  expression for the mixing of the two  massless gravitons: 
 $M^2 \simeq  3n^2 (\kappa_4^2 + \kappa_4^{\prime\,2})/16\pi^2$,  where  $M$ is the mass splitting of the gravitons, 
$\kappa_4^2,  \kappa_4^{\prime\,2}$ are the effective gravitational couplings
 of the AdS$_4$   Universes, and $n$ is the quantized charge of the gate.
 This agrees  with  earlier results based on double-trace deformations,  
 with the important difference that the effective   coupling is   quantized. 
 We argue that the  apparent non-localities of   
holographic   double-trace theories     are resolved 
by  integrating-in    the (scarce)  degrees of freedom of the gate. 
       }

\clearpage
\setcounter{page}{1}

 %%%%%%%%%%%%%%%%    

\section{Introduction}

One of the tantalizing aspects of General Relativity is the possibility of connecting disjoint Universes.
Most  of the attention has been captured by wormholes which are  pointlike contacts  between  
  Universes \cite{1,2,3}.  But one can  in principle   consider a wormbrane  or $Wp$-brane, 
 that is a bridge whose entry and exit 
  are of  spacetime dimension 
 $p+1$. In this language  the usual wormholes are  
 $W(-1)$-branes.

          When the Universes are AdS$_{d+1}$,  holographic duality offers a different perspective 
          of  such  objects   as  bridges between 
   two decoupled  $d$-dimensional field theories.
 Consistency requires that   non-traversable wormholes   correspond to pure entanglement of the
  theories \cite{Maldacena:2013xja},   
 while   traversable bridges  must  also involve  a Hamiltonian  coupling  \cite{Gao:2016bin}-\cite{vanBreukelen:2017dul}. The  generic  deformation    is 
  given by  a double-trace coupling
\bea
  {S}_{\rm int}  \sim \int  d^{p}\zeta\   {\cal O}(x(\zeta)) \,  {\cal O}^\prime(x^\prime(\zeta))\,  
K(x, x^\prime)\ , 
\eea
where  ${\cal O}, {\cal O}^\prime$ are  single-trace operators  in   the two theories,    $x(\zeta)$ and $x^\prime(\zeta)$
parametrize the  boundary submanifolds   sewed together by the coupling, and $K(x, x^\prime)$
is an interaction kernel. 
If one  insists on conformal invariance the  coupling will extend at all scales, and  the bridge will
have codimension $(d-p)$ both in the boundary and    in the  bulk.\,\footnote{In general, after taking account of the backreaction 
the bridge will be fat rather  than  delta-function localized in the transverse dimensions. Its worldvolume
can be either Euclidean or Lorentzian. }  This excludes the case $p=-1$. 
In this paper   we will focus on  the other  extreme,  
   $p=d$,  where   the entry and the exit of the bridge are  the entire AdS spacetime. 
   From a  higher-dimensional perspective on the other hand, the entry and exit  still look   like 
     localized defects.

           Double-trace deformations were introduced  in 
           \cite{Aharony:2001pa}-\cite{Berkooz:2002ug} and used   
   to  model two or more interacting  gravitons  in
\cite{Porrati:2003sa}-\cite{Kiritsis:2008at}. 
        Because  of the absence of the  
 van Dam-Veltman-Zakharov (DVZ) discontinuity in  Anti-de-Sitter  
spacetime \cite{Kogan:2000uy}\cite{Porrati:2000cp} 
 a linear combination 
of the massless gravitons  can obtain an arbitrarily-small   mass
 $M$.  An interesting feature of these double-trace models is that $M$ 
comes from  a one-loop  quantum-gravity effect. 
However, 
although double-trace deformations have been understood
as  boundary conditions in the supergravity limit \cite{Witten:2001ua}, 
their  status in  string theory
is   less clear. Their presence seems to introduce non-localities both in the target spacetime
and on the worldsheet
\cite{Aharony:2001pa}\cite{Berkooz:2002ug}. 
%Furthermore, they 
% can  destabilize the vacuum  and  make  calculations quite  challenging
%(see  e.g. \cite{Petkou:2002bb}-\cite{Aharony:2015afa}). 

%\smallskip
    
        The gates  presented in this paper  share one key feature with  these earlier  models:  
       $M$   is suppressed   by two  powers  of the effective gravitational coupling $\kappa_{d+1}$. 
        Contrary,   however, to  double-trace models, our   
         gates have  a good semiclassical  limit and    
               are  perfectly local 
          when viewed  both from   the boundary and from  the bulk.  
             The price to pay (as usual)   for   locality 
                is that  the  continuous double-trace 
        coupling must be traded for an integer charge.

\begin{figure}[t!]
  \vskip -5mm
\centering 
%\vskip 0.2 cm
\includegraphics[width=.80\textwidth,scale=0.74,clip=true]{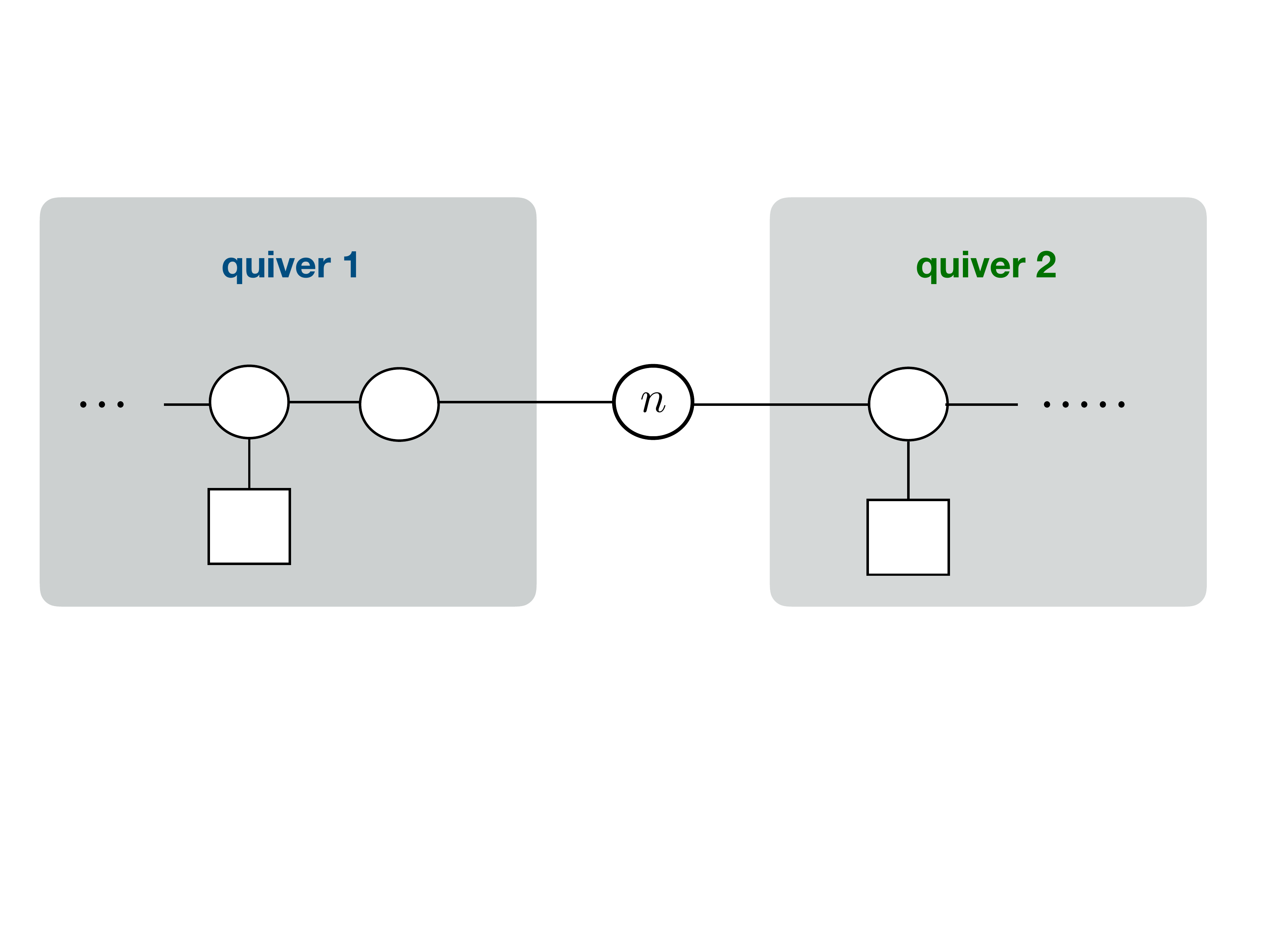}
   \vskip -19mm
   \caption{\small Two large  quivers corresponding to two large AdS Universes joined by  
   a gate  which is a low-rank gauge theory coupling via bifundamental matter to the quivers. }
\label{fig:44} 
 \end{figure}

     The   basic idea    is illustrated in   figure \ref{fig:44}\,.  One starts with  two large-quiver gauge theories
     that are dual to two large AdS$_4$   spacetimes.  The number of degrees of freedom in these quivers
     is measured by  the inverse-squared effective couplings,
      $\kappa_{4}^{-2}$ and $\kappa_{4}^{\prime\, -2}$. 
        The bridge is  then an additional 
      `messenger'  node representing  a small gauge group with  rank $n  \ll 
       \kappa_{4}^{-1}, \kappa_{4}^{\prime\, -1}$.  
   %here
       We here  consider  quivers  corresponding to
       `good'  3d ${\cal N}=4$ supersymmetric gauge theories at the origin of their Higgs or Coulomb branches
       \cite{Hanany:1996ie,Gaiotto:2008ak}          
       for which  a detailed  holographic dictionary is available   \cite{Assel:2011xz}-\cite{Bachas:2017wva}\,.
       The idea is however more general.  When $n\gg 1$  (but still   
     much smaller  than $\kappa_{4}^{-1}, \kappa_{4}^{\prime\, -1}$) 
 %here
           the bridge admits  a smooth gravitational description as a  AdS$_5\times$S$^5$ throat
      of radius $L \sim n^{1/4} $. This had been noticed already in \cite{Assel:2011xz}\cite{Assel:2012cj}. 
    Excising  the   throat is equivalent to  integrating
      out the messenger degrees of freedom leading to  two   effective  descriptions of the bridge, 
      either as the  gluing  
      of  two  AdS$_4$ spacetimes or as a multitrace 
      deformation of  the boundary theories.     In our  example  
      both these effective descriptions   are highly non-local because one  integrates out massless  fields. 
      But it should  be  obvious that this apparent non-locality is a red herring.

         To make the field-theory deformation quasi-local one may  give  mass to  the   
          hypermultiplets represented  by the two  links that  join  the 
          $U(n)$ node to the quivers.  
          The dual geometry should   now exhibit   a characteristic scale below  which
          the bridge between the Universes  disappears. 
         Taking the formal   $m\to\infty$ limit  makes the double-trace deformation 
          local, but the geometry is singular.   This explains the  tension between
          locality in field theory and in string theory. To resolve it  one must simply integrate
          back-in  the  gate  fields.

                Quivers like those of figure \ref{fig:44}  actually make   sense for any $p\leq d$  (including $p=-1$)
                and can serve as definitions of $Wp$-branes. In most cases the dual geometries are
                singular, and the problem is further complicated by infrared divergences.
                 The question of what 
                 constitutes  a `weak link'  (as opposed to a full-fledged interface) 
                  must be in particular  carefully  reexamined.   These issues
                 will be discussed elsewhere.  
                
                The plan of the present  paper is as follows:  In section \ref{2} we review some relevant features
                of the 3d ${\cal N}=4$ quiver gauge theories that we  need. We recall  in particular how the data
                for good quivers can be repackaged efficiently in 
                an ordered
                pair of Young diagrams  $(\rho, \hat\rho)$. In section \ref{3} we describe the microscopic gate
                of figure \ref{fig:44} as the rearrangement of $n$ boxes in the Young diagrams. 
                In section \ref{4} we present the dual type-IIB supergravity solutions before and after
                the construction of the  bridge.  The mixing of the gravitons due to  the introduction of 
                the gate is calculated 
                in the  semiclassical limit $1\ll n\ll \kappa_{4}^{-2}, \kappa_{4}^{\prime\, -2}$   
                in section \ref{5},  and shown to agree parametrically with the double-trace models
                of \cite{Porrati:2003sa}-\cite{Kiritsis:2008at}.  One can interpret 
                our result  as a  rule of quantization of the double-trace coupling in these models. 
                Finally, in section \ref{6} we    comment on some future directions.

%%%%%%%%%%%

\section{Partitions for  good  quivers}
\label{2}

 The field theories  of  our holographic setup are   three-dimensional ${\cal N}=4$ gauge theories 
 that can be engineered with  D3-branes suspended between D5-branes and NS5-branes
 \cite{Hanany:1996ie}. Let  $A, N, \hat N$ be respectively  the number  of these three types of  brane. 
 To define the gauge theory one must give   two ordered partitions of    $A$   in  $N$ or  $\hat N$ positive
        integers
         \bea
          A   =\  l_1 + l_2 + \cdots + l_N \ =\  \hat l_1+\hat l_2 + \cdots  \hat l_{\hat N}\ ,
         \eea
         where  $l_i\geq l_{i+1}$ and $\hat l_{\hat i}\geq \hat l_{\hat i +1}$. 
         These  describe the distribution of the D3-branes among  NS5-branes on the left
         and D5-branes on the right. 
         Equivalently, the partitions 
            define  two Young diagrams,  $\rho$ and $\hat\rho$,    both with  the same number
            $A$ of boxes. 
      The diagram $\rho$ has $l_i$ boxes in the $i^{th}$ row,  
      and $\hat\rho$ has $\hat l_{\hat j}$ boxes in the $\hat j^{\,th}$ row. 
      %It can be viewed as a directed walk
      % starting at position $l_1$ and ending after $N$ `ticks of the clock'  at position $0$.
        %At each tick of the clock the walker takes $s_i$ steps where
       % $s_1 = l_1 - l_2$,  $s_2=l_2-l_3 , \cdots$$, s_N= l_N$. 
       We label the rows of the transposed Young diagram  $\rho^T$ (i.e. the columns of $\rho$)   by
       hatted Latin letters, and the rows of the transposed Young diagrams  $\hat\rho^T$ by unhatted 
        letters. The reason for this notation will  soon be  clear. The length of the $\hat j^{th}$ row in $\rho^T$
       is   $l^{\,T}_{\hat j}$, and  the length of the $j^{th}$ row in  $\hat\rho^T$
        is  $\hat l^{\,T}_j$\,. 
       
        %  The number of boxes, $A$,  in the partition $\rho$ is also denoted by $\vert\rho\vert$.  
   %\smallskip 
 \smallskip
 
    Quivers whose gauge symmetry can be entirely  Higgsed   correspond to
    pairs   obeying the ordering condition  $\rho^T > \hat\rho$. It was
    conjectured by Gaiotto and Witten \cite{Gaiotto:2008ak} 
    that at the origin of their Higgs branch such `good theories' flow 
    to strongly-coupled supersymmetric CFTs that are irreducible with no free-field  factors. 
                We can put   the ordering  condition   in compact form by introducing the integrated row lengths
              \bea
               L_j =   \sum_{i=1}^j l_i\ \ ,  \qquad  L_{\hat j}^{\,T} =  \sum_{\hat i=1}^{\hat j}  l_{\hat i}^{\,T}\ \ ,
               \qquad   \hat L_{\hat j}  =  \sum_{\hat i=1}^{\hat j}  \hat l_{\hat i}\ \ ,  \qquad
               \hat L_{ j}^{\,T} =  \sum_{  i=1}^{ j}  \hat l_{  i}^{\,T}\ \ , 
              \eea
              which  count the total number of boxes in the first $j$ or $\hat j$ rows of the  
              corresponding diagrams.    
                  The condition $\rho^T > \hat\rho$ 
                    is  then equivalent to the following set of strict   inequalities
                   \bea\label{ordering}
                    L_{\hat j}^{\,T}  > \hat L_{\hat j} \qquad {\rm for\ all}\quad \hat j= 1, 2 \cdots,  \hat N-1 \ . 
                      \eea   
                In words, the first $\hat j$  rows of $\rho^T$    contain more 
                boxes than the first $\hat j$ non-empty
                 rows of $\hat \rho$, for all $\hat j$.  The mirror statement  $\hat\rho^T>\rho$ 
                 can be shown  to be  mathematically equivalent.

\begin{figure}[t!]
  \vskip 5mm
\centering 
%\vskip 0.2 cm
\includegraphics[width=.75\textwidth,scale=0.76,clip=true]{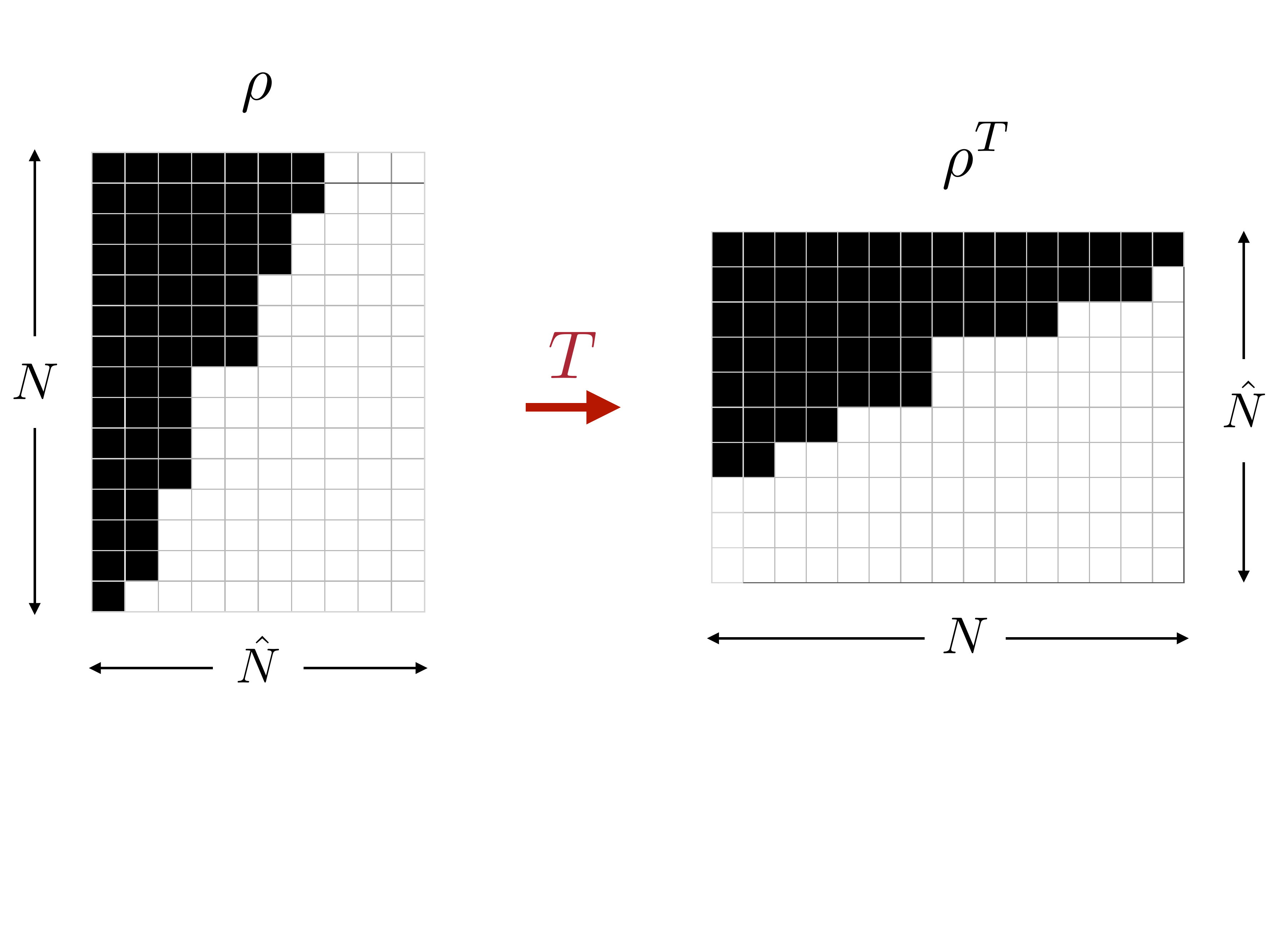}
   \vskip -21mm
   \caption{\small The  Young diagram $\rho$ and its transpose $\rho^T$
    inscribed in  their respective grids.
   %as explained in the text. Viewed as a directed walk, $\rho^T$ is obtained from $\rho$ by a
   %reflection and 90$^o$ rotation. 
   }
   
\label{fig:1} 
 \end{figure}

        The first of the above inequalities implies that  $N> \hat l_1$,
                 while its mirror statement is $\hat N>l_1$. 
             It follows that    the Young diagrams  
            $\rho^T$ and $\hat\rho$ are   contained   in a $\hat N\times  N$ grid, 
            and the diagrams $\rho$ and $\hat\rho^T$ are both contained  in a $N\times \hat N$ grid, 
             see  figure \ref{fig:1}\,.  
                    This justifies   our use of the same  labelling  for the rows of 
            $\rho^T$ and $\hat\rho$,  and also  for the rows of $\rho$ and $\hat\rho^T$. 
            When viewed as directed walks $\rho$  and $\hat\rho$ end at the lower left corner of their  
            respective grids, 
            while  the  transposed walks begin 
            at the upper right corner of their grids.
           %  Note that the  walks defined by  $\rho$  and $\hat\rho$ 
           % end at the lower left corner of their  grids, while  the  transposed walks begin 
            % at the upper right corner.  In  the transposed diagrams some of the bottom rows are 
            % necessarily empty. 

\begin{figure}[b!]
  \vskip -7.3mm
\centering 
%\vskip 0.2 cm
\includegraphics[width=.88\textwidth,scale=0.76,clip=true]{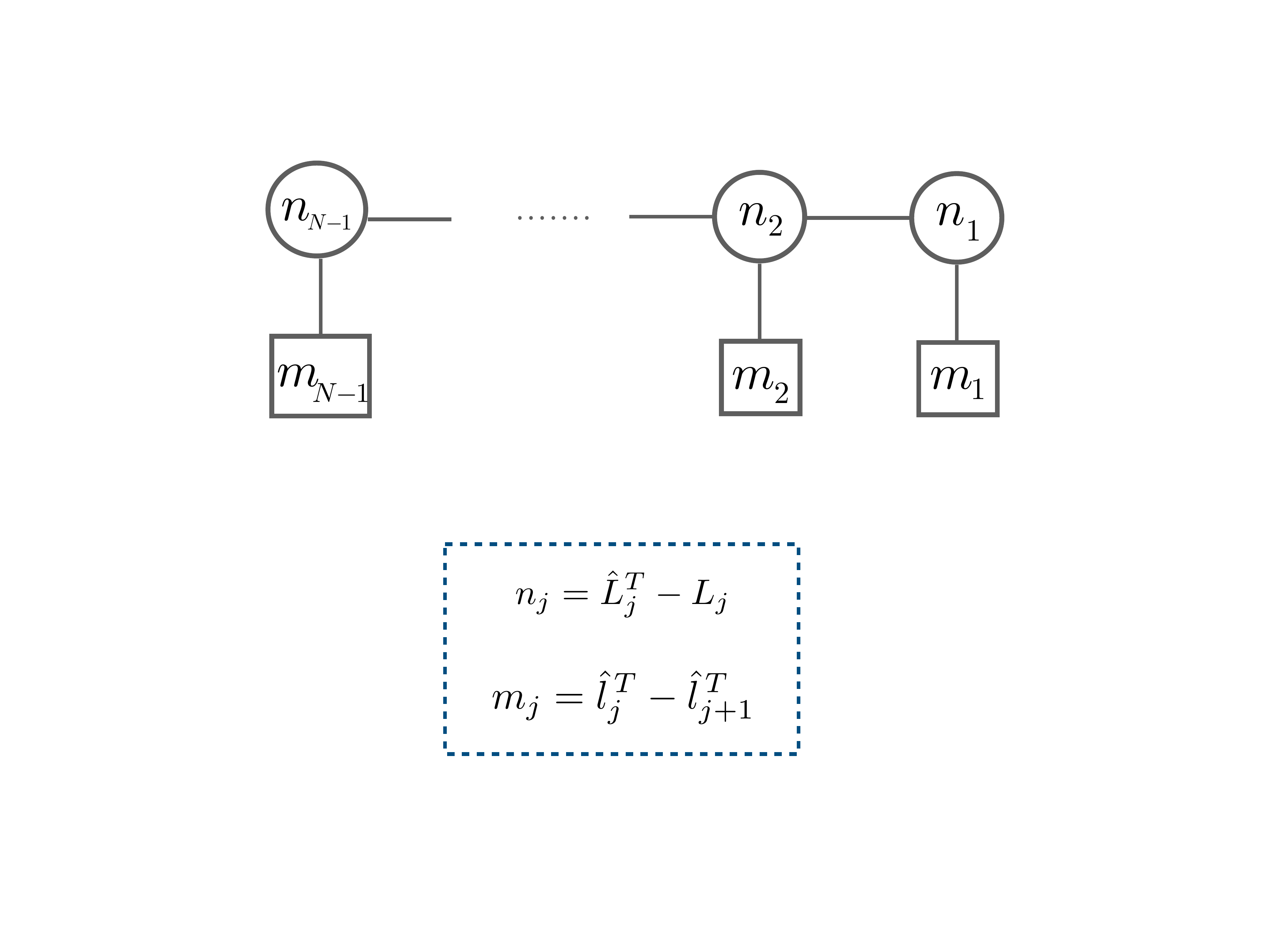}
   \vskip -14.3mm
   \caption{\small The magnetic  quiver for the ordered pair of partitions $(\rho, \hat\rho)$. 
   The gauge-group ranks $n_j$ and the flavor-group ranks $m_j$ can be expressed in terms of
   the row lengths of  $\rho$ and 
   $\hat\rho^T$. The inequality $\hat\rho^T >  \rho$ guarantees
   the positivity of all $n_j$.}
\label{fig:2} 
 \end{figure}

                    %\smallskip       
    The linear-quiver theories defined by such   partitions are called 
    $T_\rho^{\hat\rho}[{ SU}(A)] \equiv  T^\rho_{\hat\rho}[SU(A)]$  where `$\equiv$'  denotes
     mirror symmetry. 
    Their quivers are shown   in  figure \ref{fig:2}\,.    We call electric the quiver  with $\hat N- 1$ nodes (for which  the gauge group is realized on  D3-branes suspended on

\noindent      NS5-branes)  and magnetic the quiver with $N-1$ nodes (with the D3-branes suspended
on D5-branes).  
     To minimize the occurence of  hatted symbols  we choose to show here the magnetic quiver.  
      The ranks  $n_j$ and $m_j$ 
     of   the gauge  and    the flavor groups
     can be read from the row-lengths of  $\rho$ and $\hat\rho^T$ via the relations
     \bea\label{magnetic}
       n_i =  \ \hat L_j^T - L_j\ , \qquad m_j = \hat l^{\,T}_j - \hat l^{\,T}_{j+1}\qquad\qquad 
       (j= 1, \cdots , N-1)\ . 
      \eea 
       The ordering condition
             $\hat\rho^T >  \rho$ ensures
   that  all gauge-group factors have positive rank, while for the flavor groups this is automatic.
  The dual electric  quiver can be expressed
   likewise   in terms of  the  row lengths of $\rho^T$ and $\hat\rho$.

\begin{figure}[t!]
  \vskip -11mm
\centering 
 %\vskip 0.2 cm
\includegraphics[width=.74\textwidth,scale=0.70,clip=true]{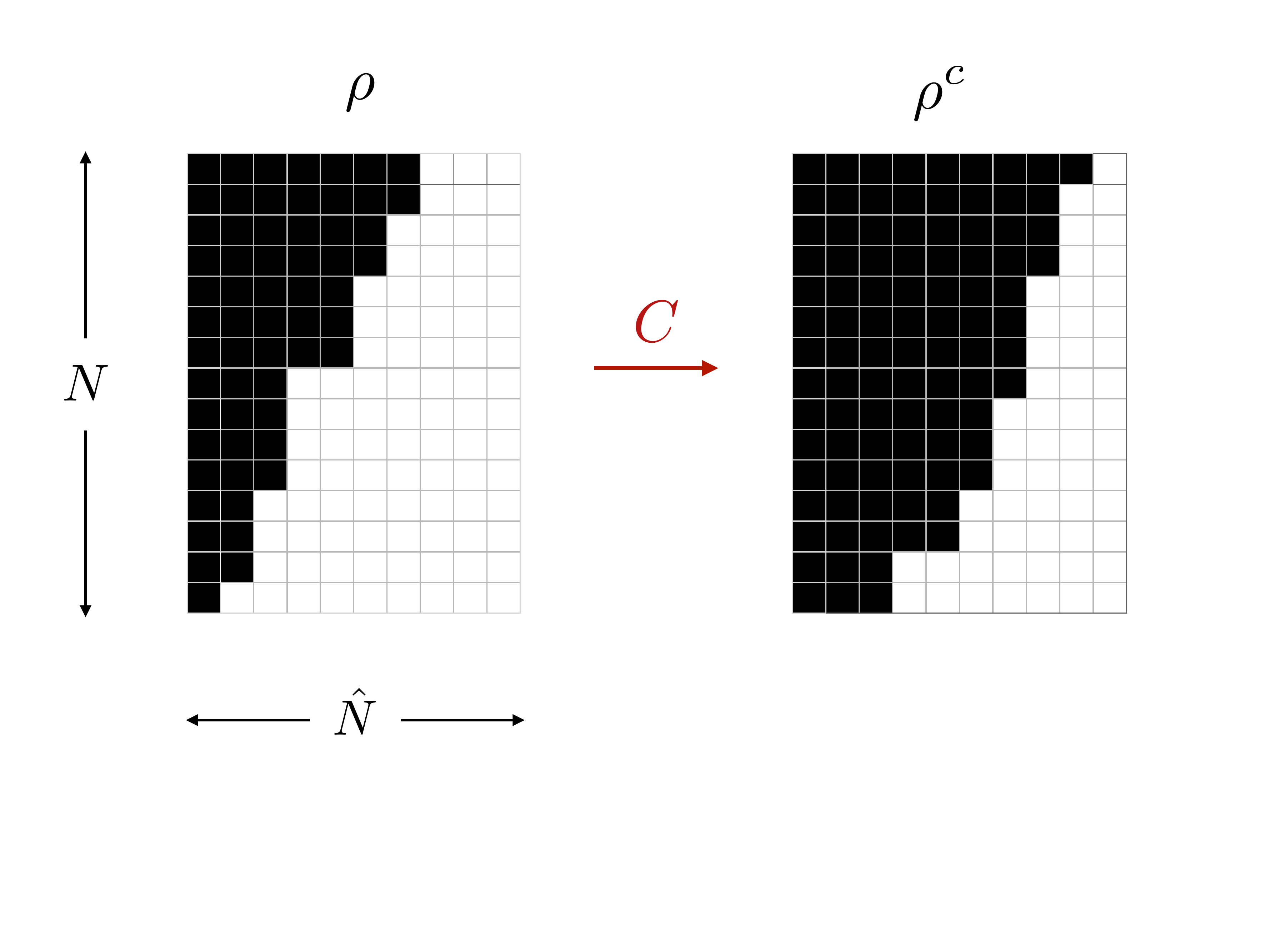}
   \vskip -18mm
   \caption{\small The operation $C$ that  replaces   $\rho$ by its complement
   inside the  $ N\times  \hat N$ grid, 
   and  $\hat\rho$ by its  complement inside the   $N\times  \hat N$
   grid. $C$  changes black to white and rotates the diagram by 180$^o$.}
\label{fig:3} 
 \end{figure}

 % \smallskip
    Besides  mirror symmetry  which exchanges $\rho$ and $\hat\rho$,  good  pairs 
     of Young diagrams  admit  one other   involution ($C$)  which replaces
      $\rho$ by its  complement $\rho^c$ inside the  $N\times \hat N$  grid, and $\hat\rho$
      by its compliment $\hat\rho^c$ inside the $\hat N\times  N$  grid, as in 
   figure \ref{fig:3}\,.  The lengths of  the rows in the transformed diagrams are    
    \bea
       l_j^c  = \hat N - l_{N-j}  \, \qquad  {\rm and} 
     \qquad   \hat l^{\,c}_{\hat j}=  N - \hat l_{\hat N -\hat j}\ . 
   \eea
  The  reader is invited to   check that   this  operation  amounts to a reflection
  of    the electric and magnetic
  quivers.  In the underlying string theory this  flips the orientation of the suspended D3-branes. 
   Since the ${\cal N}=4$ gauge  theories are  not chiral,  $C$  is   a symmetry of the 
   problem.\,\footnote{$C$ changes  $A$ to $(N\hat N-A)$, in apparent violation of the 
    D3-brane charge. It is however known that this latter is only defined modulo 
   large gauge
   transformations, and  can be shifted by  $N\hat N$. This shift changes the charge to  $-A$ 
  consistently with the fact that   $C$   reverses  the orientation  of the D3-branes.  
   }

 %%%%%%%%%%%%%%%%%%%                                                                                                          
 
 \section{Quantum gates  as  box moves}                                                                                                                                                                                
  \label{3}
                                                                                                                                                                                                                                                                                                                                                                  
 Consider now  two decoupled theories, described by the good pairs  
 $(\rho_{1}, \hat\rho_{1})$  and  $(\rho_{2}, \hat\rho_{2})$. 
 We assume that all the brane charges are large, so that the
   dual AdS$_4$ spacetimes can be described accurately by type-IIB supergravity.
   We would  like  to couple these theories  weakly,   as shown in the figure below.  `Weak' 
   means  that the node joining the quivers has a gauge group of 
    low rank $n$.  The weakest bridge has $n=1$. 
    %We will make this  statement more precise later.
     When  the  quivers in the picture are magnetic, we call  this  
    an `elementary magnetic  bridge'  between theory 1 and theory 2. 
    Since we may  join either of the two ends of each quiver, there exist  four
    different magnetic bridges  between two theories, and also four  different electric bridges.
  
\begin{figure}[t!]
  \vskip -15mm
\centering 
%\vskip 0.2 cm
\includegraphics[width=.83\textwidth,scale=0.74,clip=true]{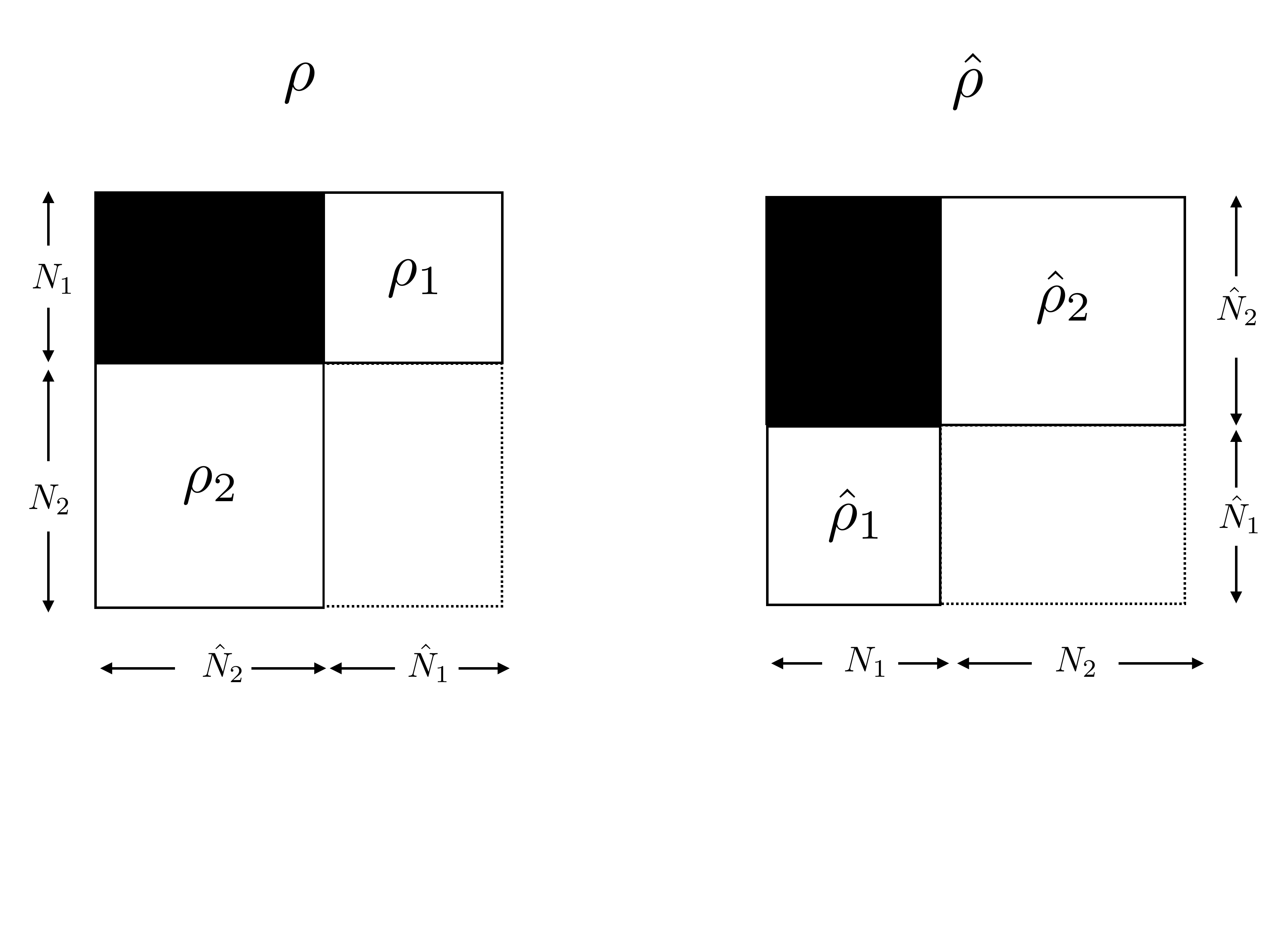}
   \vskip -22mm
   \caption{\small  The Young diagrams $(\rho, \hat\rho)$ corresponding to two decoupled theories  
   $(\rho_1, \hat\rho_1)$ and $(\rho_2, \hat\rho_2)$.
   }
\label{fig:4} 
 \end{figure}

       Before describing the bridge, let us first construct  the                    
         partition  pair $(\rho, \hat\rho)$ in  the decoupled case, $n=0$. 
  Together the two quivers have $(N_1+N_2)$  D5-branes and $(\hat N_1+\hat N_2)$
   NS5-branes, so the grid containing $\rho$ must have  dimensions  $(N_1+N_2)\times (\hat N_1+\hat N_2)$, 
   and the grid containing $\hat\rho$ must have  dimensions  $(\hat N_1+\hat N_2)\times (N_1+ N_2)$. 
    The  partitions corresponding to  the product theory are 
      shown in  figure \ref{fig:4}\,.   Their  Young diagrams 
    contain all  
     the boxes in  the black upper-left blocks of the grids,  
    and none  
    of the  boxes in  the white lower-right blocks.  The off-diagonal  blocks contain
     the diagrams of theory 1 and 2, as shown in the figure.
     
   To construct the  magnetic quiver of the composite theory  one looks  at   eqs.\,\eqref{magnetic}.
  The first $(N_1-1)$
nodes  reproduce the quiver of theory 1,  but at the next  node one finds a  gauge group of zero rank, 
$\,n_{N_1}= \hat L_{N_1}^T - L_{N_1}=0$. The  remaining nodes,  $j> N_1$, 
 reproduce the quiver  of  theory 2.  
 The fact that the bridge has $n=0$ rank means that 
 the partitions $\rho$ and $\hat\rho$  fail to obey  strict
ordering  at the $N_1^{th}$  node where the theories decouple.\,\footnote{If $n$
were negative, the corresponding node would have anti-D3 branes breaking supersymmetry.}

   It should now be clear  how   to create  a  bridge. We  must  
    crank up the rank of the $N_1^{th}$ gauge-group factor 
   by rearranging a few boxes of  these  diagrams. The rearrangement   should  restore the strict
    inequalities $\rho^T>\hat\rho$ that characterize irreducible quivers. 
To  visualize the construction of the bridge  let us assume  that the Young 
  diagrams of the original theories are 
rectangular blocks.\,\footnote{ In the quiver theories, this assumption leads to single-factor flavor
 groups. 
  In the dual type-IIB solutions,  it   corresponds  to  single  stacks of 5-brane sources of each type. }
 It  will soon become  clear   that the construction  is general and does not 
 depend on  this simplifying assumption. 
For now  take $\rho_p$ ($p=1,2)$  to be  rectangular 
$N_p\times l_{1,p}$ blocks,  and  $\hat\rho_p$   rectangular 
$\hat N_p\times \hat l_{1,p}$ blocks, where $l_{1,p}$ and $\hat l_{1,p}$ are the 
sizes of the longest rows, i.e. with our simplifying assuption of all rows.
Recall that these  lengths are bounded respectively    by 
  $\hat N_p$ and 
 $N_p$. 

\smallskip

\begin{figure}[t!]
  \vskip -11mm
\centering 
%\vskip 0.2 cm
\includegraphics[width=.88\textwidth,scale=0.74,clip=true]{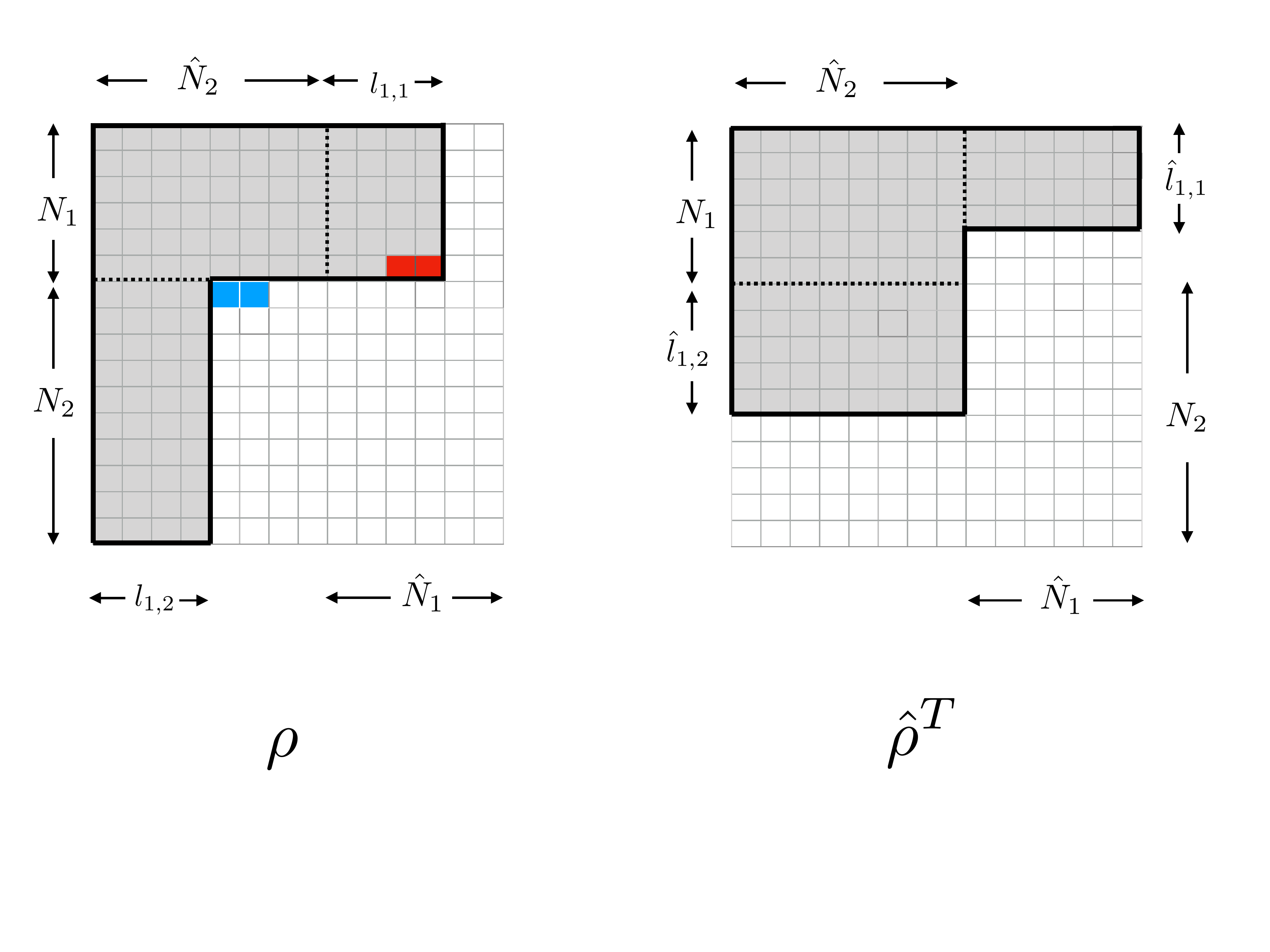}
   \vskip -18mm
   \caption{\small  The Young diagrams $\rho$  and $\hat\rho^T$ obtained by  merging two
   single-stack quivers, as discussed in the text.  A  magnetic  bridge  
   is created by moving $n$  boxes from the red to the blue positions in the diagram  $\rho$,
   while leaving $\hat\rho$ the same.  [In this  example  $N_1=  \hat N_1=6, N_2=10, \hat N_2=8$
   and $l_{1,1}=  l_{1,2}=   \hat l_{1,1}=4, \, \hat l_{1,2}= 5$.  The original diagrams contain 
   $A_1  = 24$  and $A_2 = 40$
   boxes, so for the merged diagrams  $A = 64$]\,.
 }
\label{fig:5} 
 \end{figure}

Figure \ref{fig:5} shows the   Young diagrams $\rho$ and $\hat\rho^T$ before and after the 
construction of   a  bridge.  The initial  diagrams   have the general   form  
of  figure  \ref{fig:4}\,. A magnetic   bridge 
can be constructed by   moving 
 $n $ boxes of the diagram  $\rho$
 from the $N_1^{th}$ to the $(N_1+1)^{th}$ row. This  
   increases  the rank of the  $N_1^{th}$ node from $0$ to   $n$,  leaving all other quantum numbers
 in  the magnetic quiver  unchanged, see  equations\,\eqref{magnetic}.  
 The rank of the gauge group at the connecting  node  is bounded by
 \bea
 2n \, \leq \,  \hat N_2 - l_{1,2}+ l_{1,1}\ . 
\eea
Since  the right-hand-side  is at least equal to $2$,  elementary bridges   are always allowed.
For $n>1$  there exist several rearrangements of the boxes that respect the strict ordering.
They all look indistinguishable at leading order in $n/N\hat N$ as will become clear
in the following sections. The elementary $n=1$ bridge requires the  rearrangement of a  single box
  and  can be considered   as  the   quantum of a gate.

  The reader can  easily  convince herself that the  simplifying assumption of rectangular Young
    diagrams plays no role, and that  elementary magnetic bridges between good theories always exist.
      It is also straightforward to exhibit the  electric quiver of the composite theory with  a  magnetic bridge, 
    but its detailed form  is not particularly   illuminating.  The new bridge
    has    small
   readjustments of both   gauge-group   and flavor-group ranks  at several   nodes
   of the originally-decoupled electric quivers.

 %%%%%%%%%%%%%%%%

\section{Geometry of the   gates}    
      \label{4}
        
         The   type IIB  solutions 
         dual to the ${\cal N}=4$ quiver theories were found    in  \cite{Assel:2011xz,Assel:2012cj}. 
                          The  geometry has the   warped form   (AdS$_4\times$S$^2\times \hat {\rm S}^{2}$)$\times_w \Sigma$, 
with  $\Sigma$   the infinite strip   $0\leq$Im$z\leq \pi/2$ \cite{D'Hoker:2007xy,D'Hoker:2007xz}. 
The S$^2$ fiber degenerates at the lower boundary of the strip and the $\hat{\rm S}^2$ fiber 
degenerates at the upper boundary, but  these are mere coordinate singularities.
Points where the AdS$_4$ fiber degenerates, on the other hand, are  positions of 5-brane sources. 
The D5-branes which wrap  the 2-sphere S$^2$ are 
   localized at   $z= \delta_j  + {i\pi\over 2}$ on the upper boundary of $\Sigma$, while the NS5-branes
  which 
   wrap the second sphere $\hat{\rm S}^2$ are  localized at  $z= \hat\delta_{\hat j}$
   on the lower boundary. 
   The relation of the five-brane positions   to their  linking numbers is  \cite{Assel:2011xz}
  \bea\label{basic}
    l_j =  \sum_{\hat j=1}^{\hat N}    \vartheta(\hat\delta_{\hat j} - \delta_j) \ , \qquad
    \hat l_{\hat j} = \sum_{j=1}^N  \vartheta(\hat\delta_{\hat j} - \delta_j) \ , 
   \eea 
where $\vartheta$ is the   function 
\bea
\vartheta(u) = {2\over\pi } \arctan(e^{- u})\ 
   \eea   
  which extrapolates between $1$ and $0$ as $u$ goes from   $-\infty$ and $\infty$, and 
   the five-brane singularities have been  labeled in clockwise order in order to respect   our convention that
  $\{ l_j\}$ and $\{ \hat l_{\hat j}\}$ are non-increasing sequences.  
  
    Rectangular Young diagrams
   correspond to solutions with  a single 
   stack of $N$ D5-branes all at the same position $z=\delta + {i\pi/2}$,  and 
   a single stack  of NS5-branes all at the same position $z= \hat\delta $. 
  In this case   \eqref{basic}  reduce to two equations
  \bea\label{2.2}
      { l }  =    \hat N\,  \vartheta( {\hat\delta -\delta})\ ,  \qquad
      \hat l = N \,  \vartheta( {\hat\delta -\delta}) \ ,  
  \eea
 which are related by   the conservation
  law  $Nl= \hat N \hat l$.  
  Requesting that both linking numbers be integers can make 
  this system of  equations overconstrained. 
  The  general  solution is of the form
\bea
l =      { \hat N m \over {\rm gcd}}\ , \qquad 
\hat l =      {N m \over {\rm gcd}}\ , \qquad  {\rm where}\quad 0< m<  {\rm gcd}
\eea
and ${\rm gcd}$ is the greatest common divisor of $N$ and $\hat N$. 
 If $N$ and $\hat N$  are relatively  prime  there is no solution whatsoever, 
 if    gcd$(N, \hat N)$ =  2  there is a   unique isolated solution \, 
   $m=1\Longleftrightarrow \hat\delta = \delta$ etc etc. 
 The fact that the  solutions to \eqref{2.2} depend on detailed arithmetic properties 
 of   $N$ and $\hat N$  is physically unreasonable, and 
  is   actually an artifact of the  assumption of single-stack five-branes.
 By allowing the stacks to split  one finds a large 
  number of nearby   solutions when the
 five-brane charges $N$ and $\hat N$ are large. 
   
\begin{figure}[h!]
  \vskip -20mm
\centering 
%\vskip 0.2 cm
\includegraphics[width=.84\textwidth,scale=0.74,clip=true]{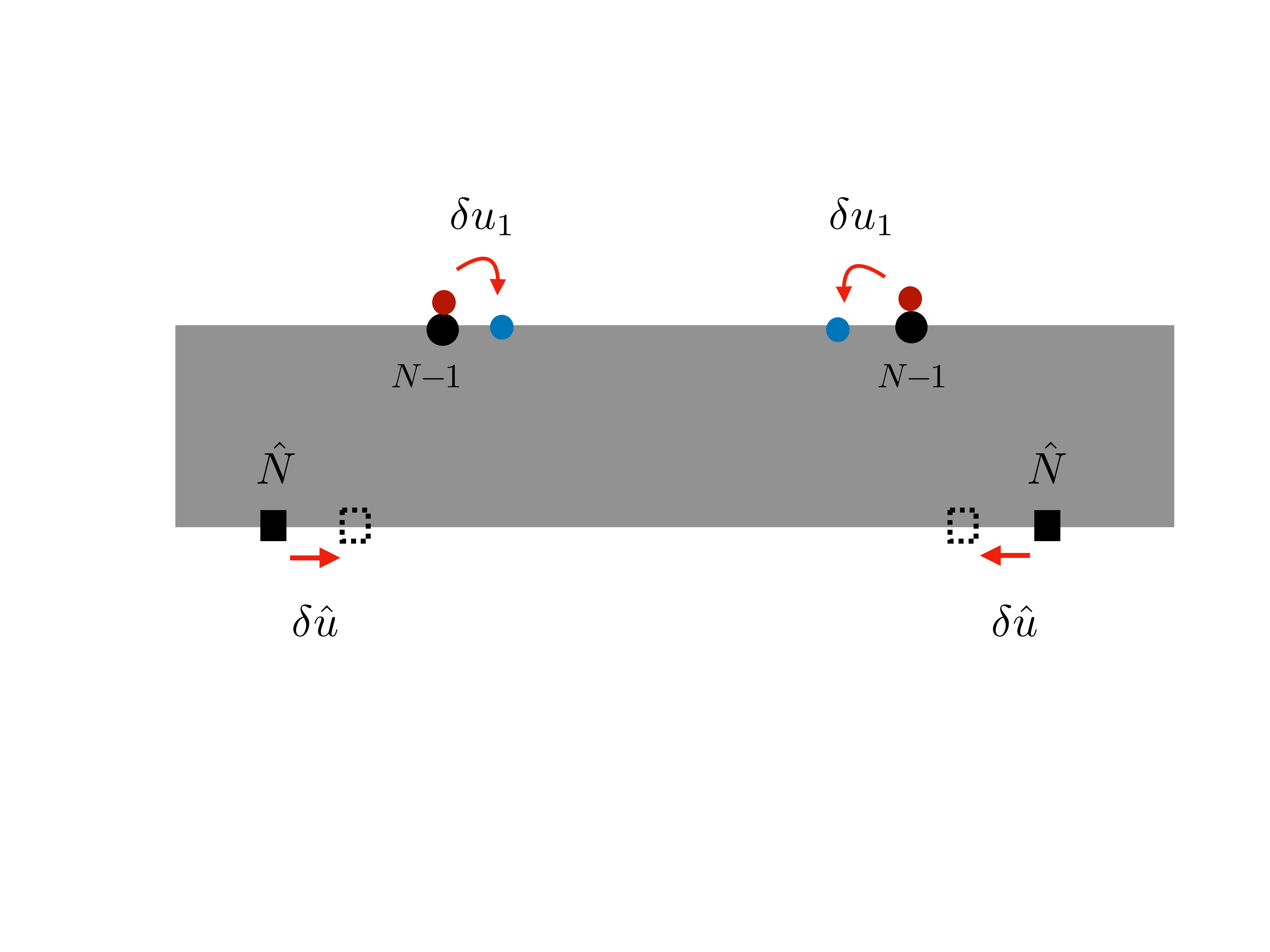}
   \vskip -20mm
   \caption{\small  The  initial geometry (upper
   part of the  figure)
  which is   dual  to the two decoupled quiver theories 
   has  singularities  separated by 
   $\xi = \infty$.  The  quantum bridge obtained by  the  rearrangement of  boxes in 
   $\rho$ shown in   figure \ref{fig:5} corresponds to  taking large but finite $\xi$ 
   and making the moves shown  in the lower part of the above 
   figure. The entire NS5-brane stacks, and one  brane  
   detached from each D5-brane stack, are  
   respectively  displaced by $\delta \hat u$ and $\delta u_1$ towards the 
   center of the strip.
 }
\label{fig:6} 
 \end{figure}

        Let us assume now that we have found a solution of \eqref{2.2} with 
       $ \delta - \hat \delta = u_0$.  To describe two   decoupled  quiver theories
       we take two copies of the above five-brane stacks with   infinite  separation 
       along the Re$z$ axis  as  in  figure \ref{fig:6}\,. To simplify the calculation, we 
       take the   symmetric 
       arrangement  shown in the figure: 
        two  stacks  of $N$  D5-branes are separated by  $\xi-u_0$,   
       and two  stacks of $\hat N$ NS5-branes are  separated by $\xi + u_0$, so that    
       the  entire configuration is invariant  under   reflection of the Re$z$- axis.

     \begin{figure}[t!]
  \vskip -18mm
\centering 
%\vskip 0.2 cm
\includegraphics[width=.84\textwidth,scale=0.74,clip=true]{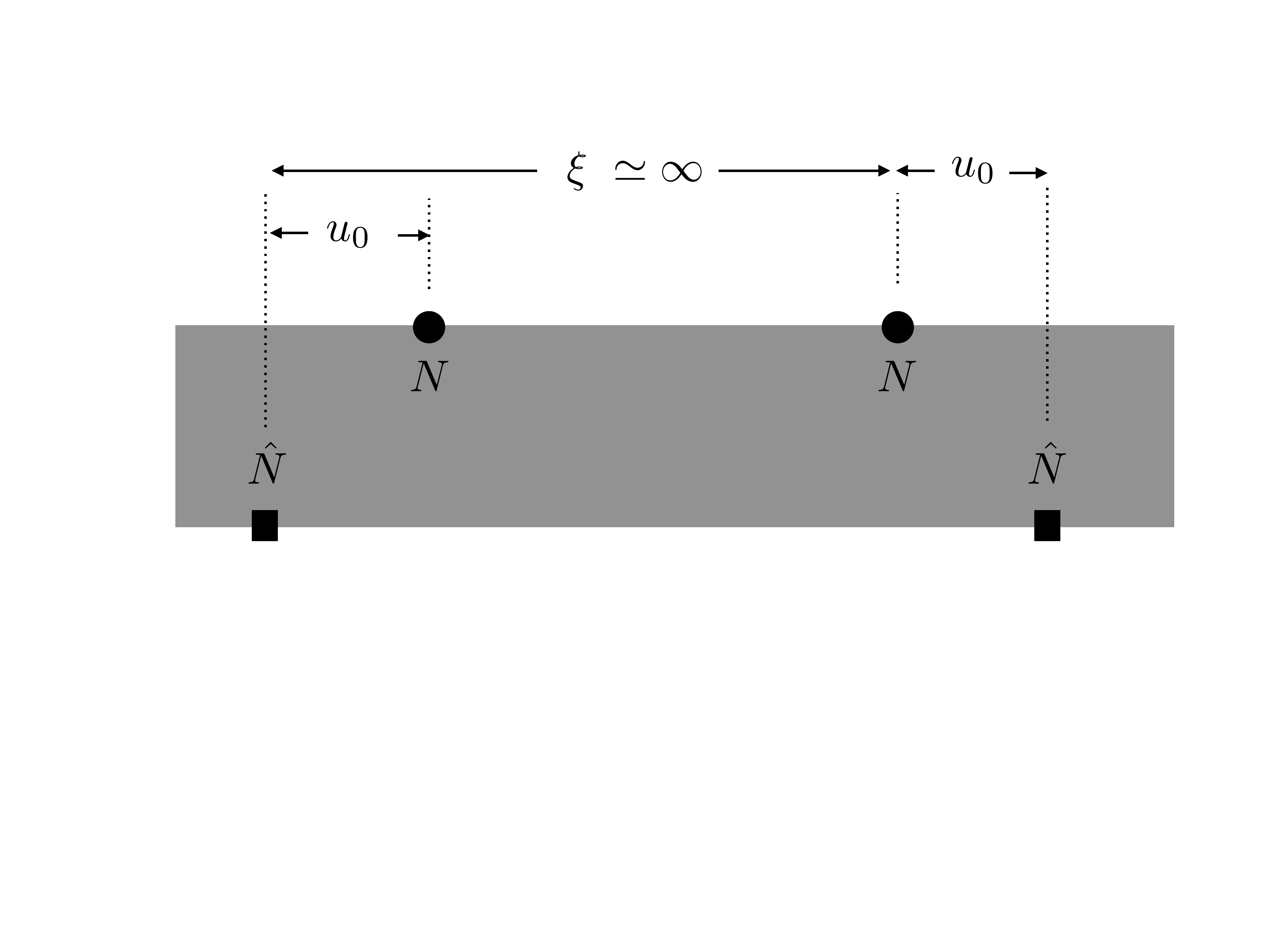}
   \vskip -23mm
    \end{figure}             
  
 %\newpage                     

   Using equations \eqref{basic}  one finds
       in the $\xi=\infty$  limit
       \bea
       l_1 = \hat N (1+ {\vartheta_0})\, , \quad l_2 =
        \hat N (1-  {\vartheta_0})\,, \quad 
       \hat l_1 = N(2-{\vartheta_0})\, , \quad \hat l_2 =  N {\vartheta_0} \ .  
        \eea 
where  
  $\vartheta_0  = {2\over\pi } \arctan(\exp(- u_0))$.  These linking numbers match   
    those of the Young diagrams for  two decoupled quivers, see  figure \ref{fig:4}\,,  if
    one   identifies 
     $N_1=N_2=N$,  
   $\hat N_1=\hat N_2= \hat N$,  and 
  $$ l_{1,1} = \hat N {\vartheta_0}\,,\quad  \hat l_{1,1} =  N {\vartheta_0}\,,\quad
    l_{1,2} = \hat N (1- {\vartheta_0})\,,\quad \hat l_{1,2} =  N (1-{\vartheta_0})\ .
    $$
    Notice  that   theory 2   is the 
  $C$-transform  of theory 1 defined in  figure \ref{fig:3}\,. This is expected  since the two theories
  are  obtained by Re$z$ reflection from  each other. 
   Of course the ${\cal N}=4$ theory is  self-conjugate, so our choice of  relative orientation
  just  indicates by which ends we chose to   join  the two decoupled quivers.

        We want now to find a new solution  obtained
         from this  initial  configuration by (i) taking  $\xi$   large but finite, and (ii) making some small  
     five-brane moves.   The  two moves that create   the elementary bridge of the previous section
               are shown in the lower part of figure \ref{fig:6}\,.   The entire   NS5-brane stacks are
               displaced by $\delta \hat u$ towards the center of the figure, while
               only a single  D5-brane   is detached  from each D5-brane stack  and 
               displaced  by $\delta u_1$ in the same direction.    
             To match   the Young diagrams of  figure \ref{fig:5}, all   linking numbers  
               except  those of the detached  D5 branes should stay the same after  these moves, while   the
                detached D5-branes should  transfer $n$ units of linking number to each other.                 
                                                                            This gives three equations for the three unknown parameters
                                                                            ($\xi, \delta \hat u$ and $\delta u_1$) 
                                                                            of the new solution
   \bea\nonumber 
   -{\hat N  } {(\delta u_1 - \delta \hat u) } \sin \pi \vartheta_0  \simeq  - \pi n\ ,\qquad
     {\hat N  } {\delta \hat u }\sin \pi \vartheta_0  - {2 \hat N } e^{-\xi}\simeq 0\ , 
    \eea      
    \bea
     {N   }  {\delta \hat u  }\sin \pi \vartheta_0  -   { \delta u_1  } \sin \pi \vartheta_0
       +  {2   N } e^{-\xi} \simeq 0\ , 
                 \eea             
      where we have neglected terms that are subleading in the limit   $N, \hat N \gg n$. 
      The solution of these leading-order equations   is  
   \bea\label{xi}
   e^{-\xi} \simeq {\pi n\over 4N\hat N}\,,  \qquad  
  \delta \hat u  \simeq  { \pi n \over  2 N \hat N  \sin \pi \vartheta_0 }  \, , 
   \qquad \delta u_1 \simeq {\pi n\over \hat N  \sin \pi \vartheta_0 }  \, . 
\eea                            
    Note that  all displacements  are  proportional to the rank $n$ of the additional gauge group
    in the magnetic quiver, and that 
         the displacement of the detached D5-branes is parametrically larger than
    that  of the NS5-branes in the large $N, \hat N$ limit.                              
     \smallskip

                The metric  of the ten-dimensional type-IIB solution  is  \cite{D'Hoker:2007xy,D'Hoker:2007xz}
   \bea\label{first}
{4\over \alpha^\prime}\, 
ds^2 = \rho_4^2\, ds^2_{\rm AdS} + \rho_1^2\, ds^2_{{(1)}} +
\rho_2^2\, ds^2_{ {(2)}} +
4 \rho^2\, dzd\bar{z} \ , 
\eea
\vskip 1mm  
\noindent where  
$ds^2_{ {(i)}} = d\vartheta_i^2 + \sin\vartheta_i^2
d\varphi_i^2\ $ are the metrics of the unit-radius
2-spheres, $ds^2_{\rm AdS}$  is the metric of the unit-radius AdS$_4$ spacetime, 
$\alpha^\prime$ is the Regge slope parameter, 
and the  four scale  factors are given by
\bea 
\rho_4^8 = 16\, {{ \cal U}_1  {\cal U}_2\over  W^2} \ ,   \quad 
\rho_1^8 = 16 h_1^8 \,{ {\cal U}_2 W^2 \over {\cal U}_1^3} \ ,
 \quad
\rho_2^8 = 16 h_2^8\, { {\cal U}_1 W^2 \over {\cal U}_2^3} \ , \quad
%\eea
 %\bea
\rho^8 = {{ \cal U}_1  {\cal U}_2 W^2 \over h_1^4 h_2^4} \,  . 
 \eea
 In the above expressions  
   \bea\label{a1}
W = \partial_z \partial_{\bar{z}} (h_1h_2) \ , \qquad  
\ { \cal U}_i = 2 h_1h_2|\partial_z h_i|^2 - h_i^2 W\ , 
   \eea                                                                    
  and $h_1$,  $h_2$
 are   
harmonic functions on the $z$-strip obtained by summing, respectively,  over the D5-brane 
and the NS5-brane singularities.
For the configuration of figure \ref{fig:6} these harmonic functions  read:   \cite{Assel:2011xz} 
                \bea \label{h1}\nonumber
 h_1 =   &&\hskip -5mm   -   (N-1)  \log    \tanh \left(   
 {{i\pi\over 4} - {z\over 2}   +  {\xi -  u_0\over 4}}   \right) 
 -   (N-1)  \log \tanh \left(   
  {i\pi\over 4} -  {z\over 2}  -  {\xi - u_0\over 4}  \right)   \
 % \  + \ c.c.  
 \\   \nonumber
  && \, \\ \nonumber &&  -   \log   \tanh \left(   
 {{i\pi\over 4} - {z\over 2}   +  {\xi -  u_0\over 4}}  - {\delta u_1\over 2}   \right) 
  -   \log  \tanh \left(   
  {i\pi\over 4} -  {z\over 2}  -  {\xi - u_0\over 4} +  {\delta u_1\over 2}  \right)    \  + \ c.c.\ , 
\eea            
                                             \bea \label{h2} 
 h_2 =   - \hat N  \log    \tanh \left(   
 { {z\over 2}  -  {\xi + u_0\over 4} + {\delta\hat u\over 2} }   \right)   - \hat N  \log
 \tanh \left(   
 { {z\over 2} +  {\xi + u_0\over 4}- {\delta\hat u\over 2} }   \right)   \
  \  + \ c.c. \ . 
\eea  
The  solutions also have a non-trivial dilaton
\bea
e^{\Phi} = \left( { {\cal U}_2\over  {\cal U}_1}  \right)^{1/4}\ , 
\eea
 as well as  3-form and 5-form backgrounds that we will not need.

Setting $z=x+iy$ and expanding these harmonic functions near the center of the strip 
($\vert x\vert \ll \xi$)  gives   after a little calculation                                           
  \bea\label{approx}
   h_1 \simeq    8N e^{-\xi/2} \cosh x \sin y \  ,  \qquad
   h_2 \simeq  8\hat N e^{-\xi/2} \cosh x \cos y\,\ ,  
 \eea
 where we  dropped terms of order $O( e^{-3\vert \xi- x\vert /2})$ which are subleading
 in the $\xi\to\infty$ limit. 
     Plugging these expansions  in  \eqref{first}-\eqref{a1} gives   the 
    AdS$_5\times$S$^5$ metric expressed  as an AdS$_4$ foliation over $x$.   
    The  radius $L$ and the  constant dilaton  $\Phi_0$ read 
   \bea
  L^4 =   4 \pi  \alpha^{\prime\,2}\,   n\ , \qquad    e^{\Phi_0} = \left( { \hat N \over  N} \right)^{1/4}\ .  
\eea
  As  expected, the radius  only depends on  the number $n$  of D3-branes that created  the  AdS$_5$ 
        throat/bridge.    
      We are here working in units $g_s=1$
      where  the NS5-branes and the D5-branes have equal tension.
       The AdS$_5$ throat  does not of course extend out to  infinity, it
         is cut off at $x\sim \pm \xi /2$ where the AdS$_5$ boundary  is capped.

  %%%%%%%%%%%%%
    
  \section{Mixing of the gravitons }   
        \label{5}

       We will compute   the mixing of the gravitons in the regime $1 \ll n \ll N\hat N$,  in which 
       the bridge is thin compared to the AdS spacetimes on either side, 
       but supergravity can   be trusted.      The general expression for the   spectrum of
       spin-2 excitations in  any warped supergravity background was given in \cite{Bachas:2011xa}.  
       The relevant eigenvalue  problem  depends only on
       the metric $g_{(6)}$ of the compact space  ${\cal M}_6$,  and on the warp factor $e^{A}\equiv \rho_4$.  The  mass-squared   operator and the norm of  wavefunctions read 
        \bea
       % {\alpha^\prime\over 4}
         M^2 =   -{e^{-2A}\over \sqrt{g_{(6)}}} \,\partial_a \sqrt{g_{(6)}}\, e^{4A} g^{ab} \partial_b  \ , \qquad 
       \vert\hskip -0.3mm \vert\psi\vert\hskip -0.3mm \vert^{\,2} \, = \,  
  \int_{{\cal M}_6}  \hskip -1.5mm \sqrt{g_{(6)}}\,  e^{2A} \, \psi^*\psi \ ,  
           \eea   
         where $\psi$ is a scalar wavefunction on ${\cal M}_6$. 
         Here $M^2$ is the dimensionless mass, which is the eigenvalue of the Lichnerowicz Laplacian
         (the spin-2 wave operator) on the  unit-radius AdS$_4$ spacetime. 
           It is related to the scaling dimension of the dual operator by the well-known formula
             $\Delta(\Delta -3)  = M^2$.  
                     For  the  case at hand ${\cal M}_6 =
         ({\rm S}^2\times    {\rm S}^{2\,\prime})\times_w\Sigma$,  and using our expressions
         for the scale factors we find\,\footnote{Note  that the metric on the strip is  $4\rho^2 dz d\bar z$,  
         giving   the volume element
 $\sqrt{g}\, d^2z = 4\rho^2 dxdy$
 and the scalar wave operator
 $g^{ab}\partial_a\partial_b =   \rho^{-2}\bar \partial_{\bar z}\partial_z$. 
 The factor  $(4\pi)^2$  comes from  the   volume of the  2-spheres.}
         \bea
         \vert\hskip -0.3mm \vert\psi\vert\hskip -0.3mm \vert^{\,2}  
 \,  =\,   (4\pi)^2 \hskip -0.8mm \int_\Sigma dxdy \,  (4 \rho^2 \rho_1^2\rho_2^2\rho_4^2)\, \vert\psi\vert^2
\, =\,  2^9  \pi^2 \hskip -0.8mm \int_\Sigma  dxdy \,  
h_1h_2\, \vert  \bar \partial\partial (h_1h_2)\vert \, 
\vert\psi\vert^2 \ , 
                \eea       
    \bea\label{mass}
     \langle \psi\vert M^2\vert \psi\rangle = 
    (4\pi)^2 \hskip -0.8mm \int_\Sigma  dxdy \,  ( 4 \rho_1^2\rho_2^2\rho_4^4)\,  
     (\bar\partial \psi^*)\partial_z \psi \, =\,  2^{10}  \pi^2 \hskip -0.8mm \int_\Sigma  dxdy \, 
     (h_1h_2)^2\,\vert \partial_z\psi\vert^2\ . 
                \eea    
    These expressions are valid  for any  of the AdS$_4$ solutions  in    \cite{Assel:2011xz,Assel:2012cj},  
     we will now
    specialize to  the nearly-factorized configurations \eqref{h2}.

              Consider first the decoupling limit $\xi\to\infty$. 
           Each AdS$_4$ spacetime  has a  massless graviton  with constant wavefunction $\psi_0$, 
           and a tower of massive excitations with $M \sim O(1)$.  
           The normalized wavefunction of the massless graviton  is
             \bea
               \psi_0  =   \, {  V}_6^{-1/2}
               \ \quad {\rm with}\quad { V}_6 = \,  2^9  \pi^2 \hskip -0.8mm \int_\Sigma  dxdy \,  
               {h_1 }{h_2 }\, 
               \Bigl\vert \partial\bar\partial ({h_1 }{h_2 })\Bigr\vert \, :=\,  (N\hat N)^2 v_6
               \ .  
              \eea
           Here $v_6$ is a number $ \sim O(1)$ that depends on the details of each 
             decoupled   theory, and whose precise value is not important.  
           It can be computed by keeping in  $h_1,h_2$
        only the  five-branes   near $x\sim \xi/2$ for the theory on the right of the bridge, or only 
           those   near $x\sim -\xi/2$ for the theory on the left.  In the  example the
            two theories are   identical. 
              
              It is useful to  express  this  compactification volume in terms of  an  effective four-dimensional
  gravitational coupling.  Following ref.\,\cite{Assel:2012cp} one   defines 
   a consistent truncation to   four-dimensional gravity
    with effective action $S_{\rm eff} = -{(1/2\kappa_4^2)} \int d^4x \sqrt{g_{(4)}}\, (R_{(4)} + 6) 
  $ which admits  the  unit-radius AdS$_4$  as solution. 
  The relation of $\kappa_4$ to $V_6$ is 
   \bea\label{kappa}
    \kappa_4^2 \, =\,  \kappa_{10}^2\, V_6^{-1} ({\alpha^\prime\over 4})^{-4} \ ,   \qquad {\rm where}  \quad   2\kappa_{10}^2 = (2\pi)^7 (\alpha^\prime)^4\   
  \eea
 is the type-IIB gravitational coupling. 
   This parametrization is particularly  convenient  when comparing the 
    on-shell supergravity action  with    the free energy
             of the  quiver gauge theory on the 3-sphere        
              \cite{Assel:2012cp}.\,\footnote{When comparing  to \cite{Assel:2012cp} and 
             to earlier references, note that  we have here  rescaled  the harmonic functions by $\alpha^\prime/4$,
             so that the coefficients of the $\log\tanh$ contributions are  integer.  
              }   
              % We  will   use it to express the final result.
                        
 %\smallskip
 
                                   Let us consider next  the configuration  with a bridge.   The two previously massless gravitons 
    will now mix, so that the graviton  with constant wavefunction $\psi_0$ remains massless, while the
   orthogonal combination    $\psi_1$ obtains a small mass. 
            To find these new  wavefunctions,  note  that the AdS$_5\times$S$^5$ bridge makes a
             parametrically-small contribution to the compactification  volume.  Indeed, cutting off the throat at
             $x=\pm x_0$ we  find 
        \bea
        {\rm Volume}_{\rm (throat)}\  \sim\   L^8 \int_{-x_0}^{x_0} \cosh^4x\,  dx \, \sim\, n^2 e^{4x_0}\ , 
        \eea
       which should be compared to the volume of the five-brane regions $\sim (N\hat N)^2$.  
       From   \eqref{xi} one sees that the two volumes are of the same order
      if   $x_0 \simeq \xi/2$, i.e. when the   AdS$_5$ cutoff reaches the five-brane 
      regions, as should be expected.   Here we take $\xi/2 \gg x_0 \gg 1$ 
      so that the   throat volume stays  parametrically small and can be ignored.  
      The two wavefunctions at this leading order are then given by
      \bea
      \psi_0 \simeq  (2V_6)^{-1/2}\ ,  \qquad 
      \psi_1 \ \simeq \   \begin{cases}
 (2V_6)^{-1/2} \qquad \, {\rm for}\quad x > x_0\,  , \\
\, \ \psi_1(x)  \qquad\quad \   {\rm for}\ \,   -  x_0 <x< x_0\, ,    \\
 - (2V_6)^{-1/2} \quad \  {\rm for}\quad x < -  x_0\ . 
\end{cases}
      \eea
   Here $\psi_1(x)$ is an interpolating function in the throat region which must be chosen 
   so as to minimize the mass.  Note that under reflection $x\to -x$,  $\psi_0$ is  even  and $\psi_1$ is odd 
   as in   the  double-well potential of quantum mechanics. 
       
        From \eqref{mass} it follows that the only contribution to the mass 
        of the $\psi_1$ state   comes  from the
        throat region where   the geometry is  
        approximately AdS$_5\times$S$ ^5$,
      $$L^{-2} ds^2\  \simeq\    dx^2 + \cosh^2x \,  ds^2({\rm AdS}_4)
        + ds^2({\rm S}^5) \ . 
        $$
      The function $\psi_1$ that minimizes  the mass  in this cut-off AdS$_5$ throat   is a solution to the differential equation  
      \bea
      {d\over dx}\bigl(\cosh^4x\, {d\psi_1\over dx}\bigr)= 0\ \ \ \Longrightarrow\ \ \ 
      \psi_1(x) \simeq {3\over 2}   \bigl(\tanh x -  {1\over 3} \tanh^3 x\bigr) (2V_6)^{-1/2} \ .  
      \eea
       In infinite AdS$_5$ spacetime this would have been  a non-normalizable solution, but in our 
       capped off  geometry it is  normalized by imposing   a smooth interpolation  between  
       the two asymptotic  values $\pm  (2V_6)^{-1/2}$. 
          Inserting this wavefunction in \eqref{mass}  and using the harmonic functions  \eqref{approx}  
    leads to  the following expression for the  mass 
  \bea
   \langle \psi_1\vert  M^2\vert \psi_1 \rangle \simeq 
  2^{16} \pi^3 (N\hat N e^{-\xi})^2\hskip -1.2mm  \int_{-x_0}^{x_0}\hskip -1.2mm
   dx \,\cosh^4x ({d\psi_1\over dx})^2
  \simeq 2^{16} \pi^3 (N\hat N e^{-\xi})^2 \times  {3\over 2V_6}\ . 
            \eea          
     Using finally \eqref{xi} and the relation  \eqref{kappa} of $V_6$ to  the effective gravitational coupling
     we arrive at  the   main result  of this paper: 
      \bea\label{main}
       M^2 \, = \,    {3  \over 8\pi^2}\, {\kappa_4^2}  \, n^2  \qquad  (n=1,2, \cdots)   \ . 
      \eea
     {If one restores the  AdS$_4$ radius  $R$
     in this formula, 
     one finds  $M^2 =  (3G_N/\pi R^4)\, n^2$,  where $G_N$ is the four-dimensional Newton's constant.}
 
                \smallskip
               It is  straightforward to extend  this calculation to the case                
    of   a bridge connecting   AdS$_4$ 
         Universes of unequal  size.  
           The properly normalized wavefunction orthogonal to the
         massless graviton  in this case reads
   \bea\nonumber\label{ggg}  
     {\cal N}^{-1}    \psi_1(x) \ \simeq \   
\, \ {3\over 4}(V_6^\prime+V_6)  (\tanh x - {1\over 3}\tanh^3x) +  {1\over 2}(V_6^\prime -V_6) \ ,   
  \eea
  \vskip -7mm
 \bea
   {\rm where} \qquad  {\cal N}^{-1} =   \sqrt{V_6V_6^\prime(V_6+V_6^\prime)}\   
\eea
and $V_6^\prime$ $  (V_6)$ is  the compactification  volume of the Universe  on the left  (right) side
of the bridge. 
Note that this wavefunction  extrapolates between ${\cal N}\, V_6^\prime$ at $x\to \infty$,  
and $- {\cal N}\,  V_6$ at $x\to -\infty$. 
        Inserting it  in the expression for the mass gives 
        \bea\label{mm}
     M^2 =   {3   \over 16 \pi^2} (\kappa_4^2 + \kappa_4^{\prime\, 2} )\, n^2  \ , 
      \eea    
      where $\kappa_4$ and
     $\kappa_4^\prime$ are the effective gravitational couplings for 
       the two theories.          
     For identical Universes this reduces to \eqref{main}.  Note that for unequal Universes  the
     mixing is dominated by the smaller Universe whose effective Newton's constant is the strongest.
                        
%%%%%%%%%%%%%%  
  
   \section{Concluding  Remarks}
   \label{6}

      We may  compare our   result for graviton mixing  with the one obtained by   Aharony et al \cite{Aharony:2006hz}
      in the double-trace deformation model. 
       Their  field theory calculation gives a mass that  
        depends on a 
      continuous double-trace coupling $h$ (in which we  reabsorbed  numerical factors) 
      and on the central
      charges of the two theories via the combination
      \bea
       M^2 =     h^2 ({1\over c_1} + {1\over c_2})\ . 
      \eea
     This is of the same form as \eqref{mm} if one notes   that the
    central charges $c_1,c_2$,   defined as the coefficients in the
       two-point function of the energy-momentum
     tensors, can  be identified with $\kappa_4^{-2}$ and $\kappa_4^{\prime\, -2}$. 
      The  important difference is that in our model $h$ is quantized. It would be interesting to
      see if  
      this quantization rule can be also found by studying RG flows in the space of double-trace coupling.
       Note that if  one  views  the  quantum bridge as the minimal allowed coupling 
            between two mutually-hidden sectors of a theory,  the  quantization of   
            charge  ensures that the mixing   cannot be  weaker 
             than $\sim \kappa_4\kappa_4^\prime$, in harmony  with the general   spirit of the  weak
             gravity conjecture.    
   
            To an observer in Universe 1 the gate    looks like  
          a D3-brane with AdS  worldvolume.   By conservation of five-form flux, the
          exit looks like  an anti-D3 brane in Universe 2.   Since the two Universes are
          invariant under charge-conjugation, only an observer travelling through the throat
          can  compare the charges of   entry and exit. 
               
            The D3-branes are special because they have   regular extremal horizons, but other
            defects can serve as entries and exits of a bridge.  The simplest case is that of a D-instanton,
            which was identified as  a wormhole solution of type-IIB supergravity in 
            \cite{Gibbons:1995vg} and should be revisited in the light of our present discussion.
            Another  interesting question was raised by  the   recent paper  \cite{Distler:2017xba},
            which counted the number of conserved energy-momentum tensors
             in class-S theories by means of an index. It would be interesting to find a way of counting
             the number of {nearly} conserved energy-momentum tensors, i.e. of the dual spin-2
             gravitons   with mass much below the mass gap of $O(1)$.

              Finally  an obvious  question is whether,  like  D-branes, quantum gates  
              can also be described on the string   worldsheet by a modification of the rules of
              string perturbation theory.  Ideas include sigma models that flow to topological theories
              in the infrared \cite{Bachas:1992cr}, zero-size wormholes in the 2d  gravity of the worldsheet
  \cite{Aharony:2001pa},\,\footnote{This has been discussed previously   in the context of matrix models,
see  \cite{Das:1989fq}-\cite{Klebanov:1994kv} and references therein.  }
   or worldsheets with conformal interfaces
          \cite{Bachas:2008jd}. Viewing the gates as weak quiver links  may give a new breadth to
          these earlier efforts. 
    
 \medskip
 \underline{Note added}: In the extension to non-identical Universes we   implicitly assumed 
        that the bridge geometry  remains AdS$_5\times$S$^5$, 
         i.e.  that the dilaton does not vary.   This will not be the case for  arbitrary  quivers  --   
           the   generic bridge has the   Janus   geometry \cite{D'Hoker:2007xy}.
                                                                                                                                                                                                                                                                                                                                                                                                                                                                                                                                      \vskip 0.3cm

{\bf Aknowledgements}: We thank  Camille Aron, Eric Bergshoeff, Massimo 
Bianchi, Amihay Hanany, 
Gary Horowitz,  Amir Kashani-Poor, Elias Kiritsis,  Kyriakos Papadodimas,
Giuseppe Policastro and  Jan Troost for discussions.

 \newpage
 
  %%%%%%%%%%%%%%%%%%%%%%%
  %%%%%%%%%%%%%%%%%%%%%%


\begin{thebibliography}{99}     
   
   \bibitem{1}   A. Einstein and N. Rosen, 
  ``The particle problem in the General Theory of Relativity,''  Phys. Rev. {\bf 48}, 73-77 (1935).                                               
                                                                              
    \bibitem{2}   J. A. Wheeler, ``Geons,'' Phys. Rev. {\bf 97}, 511-536 (1955).                                                                                                
    
    \bibitem{3}            M. S. Morris and K. S. Thorne, ``Wormholes in spacetime and their use for
interstellar travel: A tool for teaching General Relativity,''  Am. J. Phys. {\bf 56}, 395 (1988).
                                                                                                                                                                                                                                                            
                                                                                                                                                                                                                                                                                                                                                                                          %\cite{Maldacena:2013xja}
\bibitem{Maldacena:2013xja}
  J.~Maldacena and L.~Susskind,
  ``Cool horizons for entangled black holes,''
  Fortsch.\ Phys.\  {\bf 61} (2013) 781
  %doi:10.1002/prop.201300020
  [arXiv:1306.0533 [hep-th]].
  %%CITATION = doi:10.1002/prop.201300020;%%
  %375 citations counted in INSPIRE as of 28 Nov 2017

 %\cite{Gao:2016bin}
\bibitem{Gao:2016bin}
  P.~Gao, D.~L.~Jafferis and A.~Wall,
  ``Traversable Wormholes via a Double Trace Deformation,''
  arXiv:1608.05687 [hep-th].
  %%CITATION = ARXIV:1608.05687;%%
  %21 citations counted in INSPIRE as of 28 Nov 2017  
   
 %\cite{Maldacena:2017axo}
\bibitem{Maldacena:2017axo}
  J.~Maldacena, D.~Stanford and Z.~Yang,
  ``Diving into traversable wormholes,''
  Fortsch.\ Phys.\  {\bf 65} (2017) no.5,  1700034
  %doi:10.1002/prop.201700034
  [arXiv:1704.05333 [hep-th]].
  %%CITATION = doi:10.1002/prop.201700034;%%
  %17 citations counted in INSPIRE as of 28 Nov 2017  
   
 %\cite{vanBreukelen:2017dul}
\bibitem{vanBreukelen:2017dul}
  R.~van Breukelen and K.~Papadodimas,
  ``Quantum teleportation through time-shifted AdS wormholes,''
  arXiv:1708.09370 [hep-th].
  %%CITATION = ARXIV:1708.09370;%%
  %1 citations counted in INSPIRE as of 28 Nov 2017

 %\cite{Aharony:2001pa}
\bibitem{Aharony:2001pa}
  O.~Aharony, M.~Berkooz and E.~Silverstein,
  ``Multiple trace operators and nonlocal string theories,''
  JHEP {\bf 0108} (2001) 006
  %doi:10.1088/1126-6708/2001/08/006
  [hep-th/0105309].
  %%CITATION = doi:10.1088/1126-6708/2001/08/006;%%
  %101 citations counted in INSPIRE as of 25 Nov 2017                             
                                                            
%\cite{Witten:2001ua}
\bibitem{Witten:2001ua}
  E.~Witten,
  ``Multitrace operators, boundary conditions, and AdS / CFT correspondence,''
  hep-th/0112258.
  %%CITATION = HEP-TH/0112258;%%
  %367 citations counted in INSPIRE as of 28 Nov 2017
 
 
%\cite{Berkooz:2002ug}
\bibitem{Berkooz:2002ug}
  M.~Berkooz, A.~Sever and A.~Shomer,
  ``'Double trace' deformations, boundary conditions and space-time singularities,''
  JHEP {\bf 0205} (2002) 034
  %doi:10.1088/1126-6708/2002/05/034
  [hep-th/0112264].
  %%CITATION = doi:10.1088/1126-6708/2002/05/034;%%
  %180 citations counted in INSPIRE as of 28 Nov 2017  
      
         
            
                  
%\cite{Porrati:2003sa}
\bibitem{Porrati:2003sa}
  M.~Porrati,
  ``Higgs phenomenon for the graviton in ADS space,''
  Mod.\ Phys.\ Lett.\ A {\bf 18} (2003) 1793
  %doi:10.1142/S0217732303011745
  [hep-th/0306253].
  %%CITATION = doi:10.1142/S0217732303011745;%%
  %31 citations counted in INSPIRE as of 24 Nov 2017
                                                                                                                                                                                                                                                                                                                                                                                                                                                          
                                                                                                                                                                                                                                                                                                                                                                                                                                                          
                                                                                                                                                                                                                                                                                                                                                                                                                                                          %\cite{Kiritsis:2006hy}
\bibitem{Kiritsis:2006hy}
  E.~Kiritsis,
  ``Product CFTs, gravitational cloning, massive gravitons and the space of gravitational duals,''
  JHEP {\bf 0611} (2006) 049
  %doi:10.1088/1126-6708/2006/11/049
  [hep-th/0608088].
  %%CITATION = doi:10.1088/1126-6708/2006/11/049;%%                                                                                                                                                                                                                                                                                                                                                                                                                                                          
                                                                                                                                                                                                                                                                                                                                                                                                                                                          %\cite{Aharony:2006hz}
\bibitem{Aharony:2006hz}
  O.~Aharony, A.~B.~Clark and A.~Karch,
  ``The CFT/AdS correspondence, massive gravitons and a connectivity index conjecture,''
  Phys.\ Rev.\ D {\bf 74} (2006) 086006
  %doi:10.1103/PhysRevD.74.086006
  [hep-th/0608089].
  %%CITATION = doi:10.1103/PhysRevD.74.086006;%%
  %48 citations counted in INSPIRE as of 24 Nov 2017
                                                                                                                                                                                                                                                                                                                                                                                                                                                          %\cite{Kiritsis:2008at}
\bibitem{Kiritsis:2008at}
  E.~Kiritsis and V.~Niarchos,
  ``Interacting String Multi-verses and Holographic Instabilities of Massive Gravity,''
  Nucl.\ Phys.\ B {\bf 812} (2009) 488
  %doi:10.1016/j.nuclphysb.2008.12.010
  [arXiv:0808.3410 [hep-th]].
  %%CITATION = doi:10.1016/j.nuclphysb.2008.12.010;%%
  %30 citations counted in INSPIRE as of 24 Nov 2017     
  

 
                %\cite{Kogan:2000uy}
\bibitem{Kogan:2000uy}
  I.~I.~Kogan, S.~Mouslopoulos and A.~Papazoglou,
  ``The m ---> 0 limit for massive graviton in dS(4) and AdS(4): How to circumvent the van Dam-Veltman-Zakharov discontinuity,''
  Phys.\ Lett.\ B {\bf 503} (2001) 173
  %doi:10.1016/S0370-2693(01)00209-X
  [hep-th/0011138].
  %%CITATION = doi:10.1016/S0370-2693(01)00209-X;%%
  %162 citations counted in INSPIRE as of 25 Nov 2017               

%\cite{Porrati:2000cp}
\bibitem{Porrati:2000cp}
  M.~Porrati,
  ``No van Dam-Veltman-Zakharov discontinuity in AdS space,''
  Phys.\ Lett.\ B {\bf 498} (2001) 92
  %doi:10.1016/S0370-2693(00)01380-0
  [hep-th/0011152].
  %%CITATION = doi:10.1016/S0370-2693(00)01380-0;%%
  %180 citations counted in INSPIRE as of 25 Nov 2017         

 
 
  
 %\cite{Hanany:1996ie}
\bibitem{Hanany:1996ie}
  A.~Hanany and E.~Witten,
  ``Type IIB superstrings, BPS monopoles, and three-dimensional gauge dynamics,''
  Nucl.\ Phys.\ B {\bf 492} (1997) 152
  %doi:10.1016/S0550-3213(97)00157-0, 10.1016/S0550-3213(97)80030-2
  [hep-th/9611230].
  %%CITATION = doi:10.1016/S0550-3213(97)00157-0, 10.1016/S0550-3213(97)80030-2;%%
     
      
%\cite{Gaiotto:2008ak}
\bibitem{Gaiotto:2008ak}
  D.~Gaiotto and E.~Witten,
  ``S-Duality of Boundary Conditions In N=4 Super Yang-Mills Theory,''
  Adv.\ Theor.\ Math.\ Phys.\  {\bf 13} (2009) no.3,  721
  %doi:10.4310/ATMP.2009.v13.n3.a5
  [arXiv:0807.3720 [hep-th]].
  %%CITATION = doi:10.4310/ATMP.2009.v13.n3.a5;%%
  %287 citations counted in INSPIRE as of 30 Oct 2017  
           
                                                      
 %\cite{Assel:2011xz}
\bibitem{Assel:2011xz}
  B.~Assel, C.~Bachas, J.~Estes and J.~Gomis,
``Holographic Duals of D=3 N=4 Superconformal Field Theories,''  JHEP {\bf 1108} (2011) 087
   [arXiv:1106.4253 [hep-th]].
  %%CITATION = doi:10.1007/JHEP08(2011)087;%% 
  
 %\cite{Assel:2012cj}
\bibitem{Assel:2012cj} 
  B.~Assel, C.~Bachas, J.~Estes and J.~Gomis,
  ``IIB Duals of D=3 N=4 Circular Quivers,''
  JHEP {\bf 1212}, 044 (2012)
   [arXiv:1210.2590 [hep-th]].
  %%CITATION = doi:10.1007/JHEP12(2012)044;%% 

    %\cite{Assel:2013lpa}
\bibitem{Assel:2013lpa}
  B.~Assel,
  ``Holographic Duality for three-dimensional Super-conformal Field Theories,''
  arXiv:1307.4244 [hep-th].
  %%CITATION = ARXIV:1307.4244;%%    

%\cite{Bachas:2017wva}
\bibitem{Bachas:2017wva}
  C.~Bachas, M.~Bianchi and A.~Hanany,
  ``N=2 Moduli of AdS4 vacua: A fine-print study,''
  arXiv:1711.06722 [hep-th].
  %%CITATION = ARXIV:1711.06722;%%                                                                                                                                                                                                                                                                                                                                                                                                                                                                                                                                                                                                                                                                                                                                                                                                                                                                                                                    

%%\cite{D'Hoker:2007xy}
\bibitem{D'Hoker:2007xy}
  E.~D'Hoker, J.~Estes and M.~Gutperle,
``Exact half-BPS Type IIB interface solutions. I. Local solution
and supersymmetric Janus,''
  JHEP {\bf 0706} (2007) 021
  [arXiv:0705.0022 [hep-th]].
  %%CITATION = ARXIV:0705.0022;%%
  
  %\cite{D'Hoker:2007xz}
\bibitem{D'Hoker:2007xz}
  E.~D'Hoker, J.~Estes and M.~Gutperle,
``Exact half-BPS Type IIB interface solutions. II. Flux
solutions and multi-Janus,''
  JHEP {\bf 0706} (2007) 022
  [arXiv:0705.0024 [hep-th]].
  %%CITATION = ARXIV:0705.0024;%%

%\cite{Bachas:2011xa}
\bibitem{Bachas:2011xa}
  C.~Bachas and J.~Estes,
  ``Spin-2 spectrum of defect theories,''
  JHEP {\bf 1106} (2011) 005
  %doi:10.1007/JHEP06(2011)005
  [arXiv:1103.2800 [hep-th]].
  %%CITATION = doi:10.1007/JHEP06(2011)005;%%
  %17 citations counted in INSPIRE as of 19 Nov 2017                                                                                                                                                                                  
                                                                                                                                                                                                                                                                          
                                                                                                                                                                                                                                                                          %\cite{Assel:2012cp}
\bibitem{Assel:2012cp}
  B.~Assel, J.~Estes and M.~Yamazaki,
  ``Large N Free Energy of 3d N=4 SCFTs and $AdS_4/CFT_3$,''
  JHEP {\bf 1209} (2012) 074
  %doi:10.1007/JHEP09(2012)074
  [arXiv:1206.2920 [hep-th]].
  %%CITATION = doi:10.1007/JHEP09(2012)074;%%
  %12 citations counted in INSPIRE as of 20 Nov 2017
                                                                                                                                                                                                                                                                                                                                                                                                                                                                                                                                                    
                                                                                                                                                                                                                                                                          \bibitem{Gibbons:1995vg}
  G.~W.~Gibbons, M.~B.~Green and M.~J.~Perry,
  ``Instantons and seven-branes in type IIB superstring theory,''
  Phys.\ Lett.\ B {\bf 370} (1996) 37
  %doi:10.1016/0370-2693(95)01565-5
  [hep-th/9511080].
  %%CITATION = doi:10.1016/0370-2693(95)01565-5;%%
  %261 citations counted in INSPIRE as of 30 Nov 2017
                                                                                                                                                                                                                                                                          
                                                                                                                                                                                                                                                                          
                                                                                                                                                                                                                                                                          %\cite{Distler:2017xba}
\bibitem{Distler:2017xba}
  J.~Distler, B.~Ergun and F.~Yan,
  ``Product SCFTs in Class-S,''
  arXiv:1711.04727 [hep-th].
  %%CITATION = ARXIV:1711.04727;%%
  
                                                                                                                                                                                                                                                                           


        %\cite{Bachas:1992cr}
\bibitem{Bachas:1992cr}
  C.~Bachas and P.~M.~S.~Petropoulos,
  ``Topological models on the lattice and a remark on string theory cloning,''
  Commun.\ Math.\ Phys.\  {\bf 152} (1993) 191
  %doi:10.1007/BF02097063
  [hep-th/9205031].
  %%CITATION = doi:10.1007/BF02097063;%%
  %29 citations counted in INSPIRE as of 24 Nov 2017

%here

%\cite{Das:1989fq}
\bibitem{Das:1989fq}
  S.~R.~Das, A.~Dhar, A.~M.~Sengupta and S.~R.~Wadia,
  ``New Critical Behavior in $d=0$ Large $N$ Matrix Models,''
  Mod.\ Phys.\ Lett.\ A {\bf 5} (1990) 1041.
 % doi:10.1142/S0217732390001165
  %%CITATION = doi:10.1142/S0217732390001165;%%
  %95 citations counted in INSPIRE as of 13 Dec 2017

%\cite{Korchemsky:1992tt}
\bibitem{Korchemsky:1992tt}
  G.~P.~Korchemsky,
  ``Matrix model perturbed by higher order curvature terms,''
  Mod.\ Phys.\ Lett.\ A {\bf 7} (1992) 3081
  %doi:10.1142/S0217732392002470
  [hep-th/9205014].
  %%CITATION = doi:10.1142/S0217732392002470;%%
  %28 citations counted in INSPIRE as of 13 Dec 2017


%\cite{Klebanov:1994kv}
\bibitem{Klebanov:1994kv} 
  I.~R.~Klebanov and A.~Hashimoto,
  ``Nonperturbative solution of matrix models modified by trace squared terms,''
  Nucl.\ Phys.\ B {\bf 434}, 264 (1995)
  %doi:10.1016/0550-3213(94)00518-J
  [hep-th/9409064].
  %%CITATION = doi:10.1016/0550-3213(94)00518-J;%%
  %51 citations counted in INSPIRE as of 13 Dec 2017


%\cite{Bachas:2008jd}
\bibitem{Bachas:2008jd}
  C.~P.~Bachas,
  ``On the Symmetries of Classical String Theory,''
 % doi:10.1007/978-0-387-87499-9_3
  arXiv:0808.2777 [hep-th].
  %%CITATION = doi:10.1007/978-0-387-87499-9_3;%%
  %9 citations counted in INSPIRE as of 30 Nov 2017                
                        
      
          
                 \end{thebibliography}
 \end{document}